\documentstyle[11pt,fullpage]{article}
\begin{document}
\input psfig

\begin{flushright}
UFIFT-HEP-98-26\\
  hep-th/9810066\\
\end{flushright}
\vskip 3.0cm

\renewcommand{\thefootnote}{\fnsymbol{footnote}}
\def\footnoterule{\kern-3pt \hrule width \hsize \kern2.5pt}
\def\ie{{\it i.e.\ }}
\def\eg{{\it e.g.\ }}
\def\beq{\begin{equation}}
\def\eeq{\end{equation}}
\def\beqn{\begin{eqnarray}}
\def\eeqn{\end{eqnarray}}
\pagestyle{empty}

\begin{center}
{\Large\bf Bag Model for a Link in a Closed Gluonic Chain \footnote
{Supported in part by
the Department of Energy under grant DE-FG02-97ER-41029}
}
\vskip 2.5cm
\large{Charles B. Thorn}\footnote{E-mail  address: thorn@phys.ufl.edu}
\vskip 0.5cm
{\it Institute for Fundamental Theory\\
Department of Physics, University of Florida\\
Gainesville, FL 32611}
\end{center}
\vspace{1.2cm}
\begin{center}
{\bf ABSTRACT}
\end{center}
\noindent 
The large $N_c$ limit of Yang-Mills gauge theory is the
dynamics of a closed gluonic chain, but this fact does not
obviate the inherently strong coupling nature of
the dynamical problem. However, we suggest that a single
link in such a chain might be reasonably described
in the quasi-perturbative language of gluons and
their interactions. To implement this
idea, we use the MIT bag to model the physics of
a nearest neighbor bond.

\vfill
\newpage

\def\balpha{\mbox{\boldmath$\alpha$}}
\def\bgamma{\mbox{\boldmath$\gamma$}}
\def\bsigma{\mbox{\boldmath$\sigma$}}
\def\bepsilon{\hbox{\boldmath$\epsilon$}}
\def\Nlarge{N_c\rightarrow\infty}
\def\Tr{{\rm Tr}}
\def\m@th{\mathsurround=0pt }
\def\leftrightarrowfill{$\m@th \mathord\leftarrow \mkern-6mu
 \cleaders\hbox{$\mkern-2mu \mathord- \mkern-2mu$}\hfill
 \mkern-6mu \mathord\rightarrow$}
\def\overleftrightarrow#1{\vbox{\ialign{##\crcr
     \leftrightarrowfill\crcr\noalign{\kern-1pt\nointerlineskip}
     $\hfil\displaystyle{#1}\hfil$\crcr}}}

\newcommand{\ket}[1]{|#1\rangle}
\newcommand{\bra}[1]{\langle#1|}
\newcommand{\firstket}[1]{|#1)}
\newcommand{\firstbra}[1]{(#1|}
\setcounter{footnote}{0}
\pagestyle{plain}
\pagenumbering{arabic}
The sum of (planar) diagrams which survives `t Hooft's large
$N_c$ limit of $SU(N_c)$ gauge theory \cite{thooftlargen}
can be interpreted as the quantum dynamics of a closed
chain of gluons \cite{thornfock}. The limit 
establishes a cyclic ordering of the
gluons in a color singlet state, and only nearest neighbors
on each cycle interact. As long as one stays with the
continuum theory, there are no further simplifications
possible in the asymptotically free situation: dimensional
transmutation eliminates the dimensionless coupling, trading
it for a scale, which sets the mass gap if there is one. In
that case there are no free parameters in the theory, and
its single particle spectrum is knowable only
in the context of a complete solution; it is an all or nothing affair. 
The case of ${\cal N}=4$ supersymmetric Yang-Mills theory is an exception
because that theory is superconformal and hence scale invariant.
Then the `t Hooft coupling $N_cg^2$ is a free parameter, which 
can be varied in the hope of further simplification. This is
the setting for the dramatic developments of the past year
surrounding Maldacena's proposal that the large coupling
limit is equivalent to supergravity on a five dimensional 
Anti de Sitter space-time \cite{maldacena,gubserkp,wittenholog}. 

A quantitative understanding of the single particle mass
spectrum at $N_c\to\infty$ in a
non scale-invariant gauge theory, such as QCD, remains an apparently formidible
problem, although there are suggestions and conjectures that
some form of the Maldacena conjecture might yield a solution
\cite{wittenqcdads}. Alternatively, one can introduce a cutoff
of some type, thereby temporarily adding a parameter and providing
the theory with a flexibility for further simplification, albeit
at the cost of straying from the continuum limit. 
The existing proposals for the QCD/5 dimensional string theory
connection {\it a la} Maldacena 
effectively do this \cite{grosso}. The large $N_c$ limit
of lattice gauge theory \cite{eguchik,eguchikflaw} is a more
conventional version of this approach.
Yet another is our early work with a light-front (infinite momentum frame) 
description, in which the $P^+$ carried by each gluon 
is discretized \cite{thornfishnet,browergt,thornweeparton}.
In this latter work the large coupling limit in the presence
of this cutoff leads to a string interpretation, although
the physical nature of the resulting string
is highly sensitive to the details of the cutoff procedure.

In this paper we put aside attempts at a detailed
quantitative solution of large $N_c$ gauge theory, and instead aim at an 
understanding of the qualitative features of the
gluon chain dynamics which will play a crucial role in
a more complete treatment. The overriding question,
of course, is whether the gluon interactions are sufficiently
attractive to form a bond between each nearest neighbor
pair in the chain. After that there is the issue of
the fate of the gluon spin. A potentially disastrous
outcome would be a ferromagnetic tendency for the
spins to align yielding low mass glueballs with enormous spin.
The spin-spin interaction between neighboring gluons should
have an anti-ferromagnetic character. 

With regard to the first point, it is certain that the perturbative,
renormalization group improved,
Coulomb force $N_c\alpha_s(R)/2R^2$ $\sim 3\pi/(11R^2\ln R\Lambda)$ as $R\to0$ 
is attractive between neighbors due to the color
structure imposed by the large $N_c$ limit.
However, to bond a pair of massless gluons, the coefficient of $1/R^2$
must be of order 1 (in order to balance the
gradient of the kinetic energy of order $1/R$), 
which goes beyond the validity of
perturbation theory. As $R$ increases from small values, this
coefficient grows, and it is plausible that at some critical
$R_c$ the attraction is strong enough to bond, although 
this speculation is out of perturbative control. 
However, if bonding of a neighboring pair is established, 
a chain of many gluons can then
be built from similar bonds forming between each nearest
neighbor pair. An appealing feature of this scheme is that
it does not require extrapolating the
basic perturbative gluon dynamics of any given
pair too much beyond $R_c$ where $\alpha_s=O(1)$. 

We are close here to the philosophy of the MIT bag model
of low mass hadrons \cite{chodosjjtw,chodosjjt,dgjjk}. In principle,
the complete description of a light single glueball state
should involve all scales, and there would be no {\it a priori} reason
to expect the bare gluon interactions derived from the
classical Lagrangian to have much relevance. 
However the phenomenological success of the
MIT Bag model indicates that this gloomy expectation is too pessimistic. 
In that model the  large distance effects on a given single 
hadron state are lumped into
a boundary condition on the fields which are confined to
a cavity (for simplicity, usually but not necessarily taken spherical) with
a state dependent size $R$. The fields within the cavity are then treated
perturbatively, and the color magnetic part of the interactions 
yields a qualitatively correct pattern of hyperfine splittings within
a multiplet with a fixed number of partons; for instance,
this model works very well for the 3 quark baryonic states. The actual
measured size of the splittings unfortunately requires the
coupling to be uncomfortably large: $N_c\alpha_s/\pi\approx2$.
Nonetheless the idea that confinement effects are flavor
and spin independent and that the spin dependent forces are well
described by basic tree level interactions seems a reasonable
working hypothesis for our qualitative analysis.

We therefore examine the magnetic interaction 
between two neighboring gluons on
a glueball chain. Since it will be important to compare
the size of the gluon exchange contribution to the
quartic contribution, we confine the two gluons to a
spherical cavity.\footnote{
The elements of the following calculations were drawn
from unpublished bag model calculations I did in the mid seventies
for the lightest glueball states with spin 0 and 2. Some
of my results were cited in \cite{johnsont}. Similar
calculations were independently done in \cite{jaffek}. We repeat
some of that early work here because the new context requires
adaptations which are most easily explained within a
complete self-contained account.}  This
provides an infrared cutoff in space, and also takes
rough account of the interaction of the two gluons
with the environment of the rest of the chain. The first step in 
this analysis is to find the
modes of noninteracting gluons in the cavity. This
is a standard textbook problem (see for example chapter 16
of \cite{jacksonem}).
One simply solves Maxwell's equations
with time dependence $e^{-i\omega t}$. 
\begin{eqnarray}
\nabla\times{\bf E}&=&i\omega{\bf B}\qquad
\nabla\times{\bf B}=-i\omega{\bf E}\qquad
\nabla\cdot{\bf E}=
\nabla\cdot{\bf B}=0.
\end{eqnarray} 
Then 
\begin{equation}
(-\nabla^2-\omega^2){\bf E}=(-\nabla^2-\omega^2){\bf B}=0,
\end{equation}
the solutions of which are, of course, $j_l(\omega r)Y_{lm}(\theta,\phi)$.
To solve the divergence conditions, it is convenient to divide
the modes into (TE) and (TM) modes:
\begin{eqnarray}
{\bf E}^{TE}_{lm}&=&Kj_l(\omega r){\bf L}Y_{lm}\qquad
{\bf B}^{TE}={1\over i\omega}\nabla\times{\bf E}^{TE}\nonumber\\
{\bf B}^{TM}_{lm}&=&Kj_l(\omega r){\bf L}Y_{lm}\qquad
{\bf E}^{TM}={i\over \omega}\nabla\times{\bf B}^{TM}
\end{eqnarray}
where we have taken only the regular solutions at $r=0$. 

The frequencies $\omega$ are determined by the color confining boundary
conditions of a perfect magnetic conductor: ${\bf E}_{norm}=
{\bf B}_{tang}=0$.
The lowest non-zero solution for $\omega$ occurs for the (TE) case 
at $l=1$, with the boundary condition $z_0j_1^\prime(z_0)=-j_1(z_0)$
the lowest solution of which is $\omega_0 R\equiv z_0\approx 2.7437$. 
If the cavity were an electric conductor, the boundary conditions
would have the roles of electric and magnetic fields reversed,
but we stick to the color confining conditions in this article. 
Finally, we note that we may take the vector potential of each
mode to be simply ${\bf E}/i\omega$ in which case it automatically
satisfies the Coulomb gauge condition $\nabla\cdot A=0$.
It is clear from the construction of the modes that the
angular momentum carried by each mode is just $l$.

As already mentioned the lowest frequency mode is the (TE)
case with $l=1$: 
\begin{equation}
{\bf A}^0_m= {K\over i\omega_0} j_1(\omega_0r){\bf L}Y_{1m}
={K\over i\omega_0} j_1(\omega_0r)
\sqrt{3\over4\pi}{\hat r}\times\bepsilon_m
\label{lowestmode}
\end{equation}
where the polarization vectors are $\bepsilon_1=\bepsilon_{-1}^*=i(1,i,0)/\sqrt2$
and $\bepsilon_0=(0,0,-i)$,
and $K$ is determined by the normalization condition
\begin{eqnarray}
1=\int d^3x{\bf A}^*_m\cdot{\bf A}_m&=&
2{K^2\over\omega_0^2}\int_0^R r^2dr j_1^2(\omega_0r)\nonumber\\
&=&{K^2\over\omega_0^2}{R^3}
(j_1^2(z_0)+j_0^2(z_0)-3j_0(z_0)j_1(z_0)/z_0).
\end{eqnarray}

From the above considerations we know that these modes carry 
angular momentum $j=1$.
We would like to put two gluons in this lowest mode and then study the
spin-spin interaction induced by the QCD interactions. At zeroth order the
total energy of the system is $2\omega_0=2z_0/R$.
This energy level is 9-fold degenerate
with $j=2,1,0$. We are next interested in how the interactions
resolve this degeneracy.

To find out, we set up cavity perturbation theory, in
which the zeroth order fields are precisely the  modes we have just 
discussed. We therefore expand the vector potential in
a complete set of normal modes:
\begin{equation}
{\bf A}({\bf x},t)=
\sum_n{1\over\sqrt{2\omega_n}}[a_{n} \balpha_n({\bf x})e^{-i\omega_n t}
+ a^\dagger_{n} 
\balpha^*_n({\bf x})e^{i\omega_n t}]
\label{modeexp}
\end{equation}
where $\omega_n$ is the frequency 
for the $n$th mode, and the ${\balpha_n}$ are just the vector potentials of the
normal modes, normalized according to
\begin{equation}
\int d^3x\balpha_n({\bf x})\cdot\balpha^*_m({\bf x})=\delta_{nm}.
\end{equation}
They also satisfy the completeness relation
\begin{equation}
\sum_n\balpha^k_n({\bf x})\balpha^{l*}_n({\bf y})=
(\delta_{kl}-{\nabla_k\nabla_l\over\nabla^2})
\delta({\bf x}-{\bf y}),
\end{equation}
where the delta function includes a transverse projector as required
for Coulomb gauge.
The gluon propagator has the representation
\begin{eqnarray} 
&& i\bra{0}{\rm T}[A_k(x)_\alpha^\beta A_l(y)_\gamma^\delta]\ket{0}\nonumber\\
&&\qquad\qquad\qquad=\delta_\alpha^\delta\delta_\gamma^\beta
\int {d\omega\over2\pi}
e^{-i\omega(x^0-y^0)}\sum_n\left[{\alpha^k_n({\bf x})\alpha^{l*}_n({\bf y})
\over2\omega_n(\omega_n-\omega-i\epsilon)}+
{\alpha^{k*}_n({\bf x})\alpha^{l}_n({\bf y})
\over2\omega_n(\omega_n+\omega-i\epsilon)}\right]\nonumber\\
&&\qquad\qquad\qquad=\delta_\alpha^\delta\delta_\gamma^\beta
\int {d\omega\over2\pi}
e^{-i\omega(x^0-y^0)}\sum_n{\alpha^k_n({\bf x})\alpha^{l*}_n({\bf y})
\over\omega_n^2-\omega^2-i\epsilon}
\end{eqnarray}
where we made use of the fact that 
$\sum_{\omega_n~fixed}\alpha^k_n({\bf x})\alpha^{l*}_n({\bf y})$ can
be shown to be real.

We are interested in the spin-spin interaction which has its origin
in the magnetic part of the Hamiltonian,
\begin{eqnarray}
H_{mag}&=&\int d^3x{1\over4}\Tr F_{kl}^2\nonumber\\
&=&\int d^3x{1\over4}\{\Tr(\partial_kA_l-\partial_lA_k)^2
+2ig\Tr(\partial_kA_l-\partial_lA_k)[A_k,A_l]-g^2\Tr[A_k,A_l]^2\}\nonumber\\
&=&\int d^3x\{{1\over2}\Tr(\partial_kA_l)^2
+ig\Tr\partial_kA_l[A_k,A_l]-{g^2\over2}\Tr A_kA_l[A_k,A_l]\},
\end{eqnarray}
where in the last line we have committed to Coulomb gauge
$\nabla\cdot{\bf A}=0$.
The lowest order splittings are $O(g^2)$, but of course the cubic
terms will only enter at this order in second order perturbation
theory. We first evaluate the contribution of the quartic terms
for which first order perturbation theory suffices: we simply must
evaluate the matrix elements of the quartic terms on the zeroth
order states and diagonalize the resulting matrix.

We choose the color structure of the zeroth order states corresponding
to a single link in a glueball chain:
\begin{equation}
\ket{m_1,m_2}_\alpha^\beta=
{1\over\sqrt{N_c}}(a^\dagger_{m_1}a^\dagger_{m_2})_\alpha^\beta\ket{0}.
\label{2gluestate}
\end{equation}
Note that this state is in the adjoint representation of $SU(N_c)$, as
is appropritate for a single link in a gluonic chain. Contrast
this with a typical bag calculation which is restricted to color singlets
only. The operators in this state each create a gluon in
the lowest cavity mode. We have suppressed all labels 
except spin $m_{1,2}=\pm1,0$, 
and we recall that they are matrices in color space,
$\alpha,\beta$ in (\ref{2gluestate}) being color matrix indices. The
norm of this state is unity in the limit $N_c\to\infty$ by virtue
of the factor $1/\sqrt{N_c}$. Next we evaluate the matrix element
in the limit $N_c\to\infty$:
\begin{eqnarray}
&&{}^{\alpha^\prime}_{\beta^\prime}
\bra{m_1^\prime, m_2^\prime}{-g^2\over2}
\Tr A_kA_l[A_k,A_l]\ket{m_1,m_2}_\alpha^\beta
={g^2N_c\over(2\omega_0)^2}\delta_\alpha^{\alpha^\prime}
                               \delta_{\beta^\prime}^\beta
[\alpha_{m_1^\prime}^{k*}\alpha_{m_2^\prime}^{l*}
\alpha_{m_1}^{k}\alpha_{m_2}^{l}
\nonumber\\
&&\qquad\mbox{}+\alpha_{m_1^\prime}^{k*}\alpha_{m_2^\prime}^{k*}
               \alpha_{m_1}^{l}\alpha_{m_2}^{l}
-2\alpha_{m_1^\prime}^{k*}\alpha_{m_2^\prime}^{l*}\alpha_{m_1}^{l}
\alpha_{m_2}^{k}]
\end{eqnarray}
Here the $\alpha^k_m$ are just the mode functions for the lowest
cavity mode contribution to 
$A^k$ (see (\ref{modeexp}), (\ref{lowestmode})). Inserting these explicit
expressions and integrating over space, we obtain the quartic
contribution to the energy shift matrix (stripped of color factors)
\begin{eqnarray}
\bra{m_1^\prime, m_2^\prime}\Delta_{quartic}\ket{m_1,m_2}
&=&{N_cg^2\over16\pi}{3K^4\over\omega_0^6}
\int_0^R r^2dr j_1^4(\omega_0r)\nonumber\\
&&[(\bepsilon_{m_1}\times\bepsilon_{m_1^\prime}^*)
\cdot(\bepsilon_{m_2}\times\bepsilon_{m_2^\prime}^*)
+(\bepsilon_{m_1}\times\bepsilon_{m_2})
\cdot(\bepsilon_{m_1^\prime}^*\times\bepsilon_{m_2^\prime}^*)]
\end{eqnarray}
The two terms in square brackets are commuting $9\times9$ spin matrices.
The first has eigenvalues $-1,1,2$, where a
spin 2 quintet belongs to $-1$, a spin 1 triplet belongs to 1, and
a spin 0 singlet belongs to 2. The second matrix has 
respective eigenvalues $0,2,0$, so it only serves to
raise the triplet two units yielding the pattern $-1,3,2$
for respectively spin 2, 1, and 0.
This inverted level structure means that the quartic 
interaction yields ferromagnetic spin-spin couplings.
We next turn to the contribution of the cubic terms for which we must
use second order perturbation theory. 

To do second order perturbation theory we expand the interaction
picture evolution operator to second order
\begin{equation}
{\rm T}e^{-i\int_0^Tdt^\prime H_I(t^\prime)}=I-i\int_0^Tdt^\prime H_I(t^\prime)
-{1\over2}\int_0^Tdt^\prime\int_0^Tdt^{\prime\prime}\ {\rm T}[H_I(t^\prime)
H_I(t^{\prime\prime})],
\end{equation}
and consider only the contribution of the cubic terms of $H_I$ in the
last term. To extract the second order level shift we take
matrix elements between the zeroth order states and isolate 
the coefficient of $-iT$ for $T\to\infty$. Since we are stopping
at order $g^2$, terms with higher powers of $T$ will not appear,
so this procedure gives the energy shift unambiguously.
We will also exploit the simplifications of the limit $N_c\to\infty$. 
Four of the six factors of ${\bf A}$ in the last term contract
against operators in the states, leaving the last contraction
to produce a factor of the gluon propagator
\begin{equation} 
D_{kl}({\bf x},{\bf y},x^0-y^0)\delta_\alpha^\delta
\delta_\gamma^\beta\equiv i\bra{0}{\rm T}
[A_k(x)_\alpha^\beta A_l(y)_\gamma^\delta]\ket{0}
\end{equation}
which satisfies:
\begin{equation}
(\partial_0^2-\nabla^2)D_{kl}=\delta^{tr}_{kl}(x-y).
\end{equation}
The cubic term in $H_{mag}$ can be written
\begin{equation}
H_{mag, cubic}=ig\int d^3x\Tr\partial_kA_l[A_k,A_l].
\end{equation}
Inserting this into the second order term in the evolution
operator, and contracting out a pair of gluon fields in
all possible ways leads to
\begin{equation}
-{1\over2}\int_0^Tdt^\prime\int_0^Tdt^{\prime\prime}{\rm T}[H_I(t^\prime)
H_I(t^{\prime\prime})]\to-i{g^2\over2}\int_0^T d^4x\int_0^Td^4y
D_{kl}(x,y)\Tr\ {\rm T}[J_k(x)J_l(y)],
\label{secondorder}
\end{equation}
where the current operator $J_k(x)$ is found to be
\begin{equation}
J_k(x)=A_l{\overleftrightarrow{\partial_k}}A_l
+[A_k,\nabla\cdot{\bf A}]+2\nabla A_k\cdot{\bf A}-2{\bf A}\cdot\nabla A_k.
\end{equation}
A contraction which involves $\partial_kA_l$ requires an integration by
parts, but then the surface term is proportional to $[n_kA_k,A_l]$ which
is proportional to $[A_{norm},A_{tang}]$ and so vanishes for both electric
($A_{tang}=0$) and magnetic ($A_{norm}=0$) conductor boundary conditions.

When (\ref{secondorder}) is sandwiched between two states of
the same energy, a factor of $T$ arises as $T\to\infty$ because then the
integrand is only a function of $x^0-y^0$, but $x^0, y^0$ are integrated
independently over the large interval $T$. Thus for $E_f=E_i$ we
have for the coefficient of $-iT$ as $T\to\infty$,
\begin{equation}
\bra{f}\Delta E\ket{i}={g^2\over2}\int_{-\infty}^\infty dx^0
\int d^3xd^3y
D_{kl}(x,y)|_{y^0=0}\bra{f}
\Tr\ {\rm T}[J_k(x)J_l(y)|_{y^0=0}]\ket{i}.
\end{equation}
Taking $\ket{i},\ket{f}$ to be of the form (\ref{2gluestate}), we find
that in the large $N_c$ limit there are only two essentially
different contraction patterns that survive. Note that the equivalence of
$k,x$ to $l,y$ means that we can insist that the first creation operator
in $\ket{i}$, say, contracts with one of the two fields in $J(y)$ provided we multiply
by 2. After that contraction, which can be taken in two distinct ways,
the remaining contractions are all uniquely determined when $N_c\to\infty$.
There is an annihilation graph in which both operators in $\ket{i}$
contract against the fields in $J(y)$. And there is an exchange graph
in which each current contracts with one operator in $\ket{i}$ and
one in $\ket{f}$. 

We first evaluate the exchange graph. 
Since the two gluons in our chosen states have the same mode
frequency, the currents coupling to the propagator are time independent
so the $x^0$ integral can be immediately done:
\begin{equation}
\int dx^0 D_{kl}(x,y)|_{y^0=0}=\sum_{n}{1\over 2\omega_n^2}
[\alpha^k_n({\bf x})\alpha^{l*}_n({\bf y})+
\alpha^{k*}_n({\bf x})\alpha^{l}_n({\bf y})]\equiv
G_{kl}({\bf x},{\bf y}).
\end{equation}
Clearly $G$ satisfies the static Green function equation
\begin{equation}
-\nabla^2 G_{kl}({\bf x},{\bf y})=\delta^{tr}_{kl}({\bf x}-{\bf y}).
\end{equation}

The matrix element reduces to
\begin{eqnarray}
\bra{f}\Delta E\ket{i}_{exchange}
&=&{N_c{g^2}}\delta_\alpha^{\alpha^\prime}
\delta_{\beta^\prime}^\beta\int d^3x d^3y
G_{kl}({\bf x},{\bf y}){\cal J}^2_k({\bf x}){\cal J}^1_l({\bf y})
\nonumber\\
&\equiv& {N_c{g^2}}\delta_\alpha^{\alpha^\prime}
\delta_{\beta^\prime}^\beta\int d^3x{\cal J}^2_k({\bf x}){\cal A}_k({\bf x})
\end{eqnarray}
Here ${\cal J}$ is the matrix element, stripped of color
factors, of $J$ between one gluon states:
\begin{eqnarray}
{\cal J}^1_l({\bf y})&=&
{1\over2\omega_0}(\alpha_{m_1,r}{\overleftrightarrow{\partial_l}}
\alpha^*_{m_1^\prime,r}\nonumber\\
&+&\alpha_{m_1,l}\nabla\cdot\balpha^*_{m_1^\prime}-\nabla\cdot\balpha_{m_1}
\alpha^*_{m_1^\prime,l}
+2\nabla \alpha_{m_1,l}\cdot\balpha^*_{m_1^\prime}-2\balpha_{m_1}\cdot
\nabla \alpha^*_{m_1^\prime,l})\nonumber\\
{\cal J}^2_k({\bf x})&=&{1\over2\omega_0}(\alpha^*_{m_2^\prime,r}
{\overleftrightarrow{\partial_k}}\alpha_{m_2,r}\nonumber\\
&+&\alpha^*_{m_2^\prime,k}\nabla\cdot\balpha_{m_2}-\nabla\cdot\balpha^*_{m_2^\prime}
\alpha_{m_2,k}
+2\nabla \alpha^*_{m_2^\prime,k}\cdot\balpha_{m_2}-2\balpha^*_{m_2^\prime}\cdot
\nabla\alpha_{m_2,k})
\end{eqnarray}
Notice that ${\cal J}_k^2$ is the same as $-{\cal J}_k^1$ with
the substitution $m_1,m_1^\prime\to m_2,m_2^\prime$.

We do not need a closed form expression for $G_{kl}$. Instead we can 
work with ${\cal A}_k({\bf x})\equiv \int d^2y G_{kl}{\cal J}^1_l$, which
satisfies
\begin{equation}
-\nabla^2{\cal A}_k={\cal J}^1_k=-{9K^2\over4\pi\omega_0^2}
{j_1(\omega_0r)^2\over2\omega_0 r^2}[{\bf r}\times(\bepsilon_{m_1}
\times\bepsilon_{m_1^\prime}^*)]_k,
\end{equation}
together with the magnetic conductor boundary conditions ${\cal A}_{norm}=0$
and $(\nabla\times{\cal A})_{tang}=0$. The r.h.s. doesn't require
the transverse projector because $\nabla\cdot{\cal J}^1=0$.
This equation is 
easily solved by the {\it ansatz} ${\cal A}=A(r){\bf r}\times(\bepsilon_{m_1}
\times\bepsilon_{m_1^\prime}^*)$, whence A(r) satisfies:
\begin{eqnarray}
A^{\prime\prime}+{4\over r}A^\prime={9K^2\over4\pi\omega_0^2}
{j_1(\omega_0r)^2\over2\omega_0 r^2}.
\end{eqnarray}
This last equation can be directly integrated to obtain
\begin{eqnarray}
 A(r)={3K^2\over8\pi\omega_0^3}\left[-{1\over r^3}\int_0^rr^{\prime2}dr^\prime
 j_1(\omega_0r^\prime)^2-\int_r^R{dr^{\prime}\over r^{\prime}}
 j_1(\omega_0r^\prime)^2-{1\over 2R^3}\int_0^R{r^{\prime2}dr^{\prime}}
 j_1(\omega_0r^\prime)^2\right],
\end{eqnarray}
where the integration constant is fixed by the boundary condition
$RA^\prime(R)+2A(R)=0$, which assures the vanishing of the tangential
magnetic field.
In terms of these quantities the level shift becomes
\begin{eqnarray}
\bra{f}\Delta E\ket{i}_{exchange}
&=& {N_c{g^2}}\delta_\alpha^{\alpha^\prime}
\delta_{\beta^\prime}^\beta {3K^2\over\omega_0^3}\int_0^R r^2dr
A(r)j_1(\omega_0r)^2(\bepsilon_{m_1}\times\bepsilon_{m_1^\prime}^*)
\cdot(\bepsilon_{m_2}\times\bepsilon_{m_2^\prime}^*)\nonumber\\
&=&{N_c{g^2}}\delta_\alpha^{\alpha^\prime}
\delta_{\beta^\prime}^\beta{9K^4\over8\pi\omega_0^6}\int_0^R {dr\over r}
j_1(\omega_0r)^2(\bepsilon_{m_1}\times\bepsilon_{m_1^\prime}^*)
\cdot(\bepsilon_{m_2}\times\bepsilon_{m_2^\prime}^*)\nonumber\\
&&\qquad\qquad\qquad\left[-{2 }\int_0^rr^{\prime2}dr^\prime
 j_1(\omega_0r^\prime)^2-{r^3\over 2R^3}\int_0^R{r^{\prime2}dr^{\prime}}
 j_1(\omega_0r^\prime)^2\right]
\end{eqnarray}
To simplify this integral we note that
$$\chi(z)\equiv\int_0^z z^{\prime2}dz^\prime j_1(z^\prime)^2
     ={z^2\over2}[zj_1(z)^2+zj_0(z)^2-3j_0(z)j_1(z)].
$$
Putting $z=z_0$ gives the integral required to determine 
the normalization constant $2(K^2/\omega_0^2)\chi(z_0)=\omega_0^3$.
Numerically $\chi(z_0)\approx1.1385$. Stripping off the color
factors, we have the level shift matrix for the exchange graph:
\begin{eqnarray}
\bra{m_1^\prime, m_2^\prime}\Delta_{exchange}\ket{m_1,m_2}
&=&{N_c{g^2}\over16\pi}{9\omega_0\over\chi(z_0)^2}
\left[-\int_0^{z_0}{dz\over z}
j_1(z)^2\chi(z)-{\chi(z_0)^2\over 4z_0^3}\right]\nonumber\\
&&\qquad\qquad\qquad(\bepsilon_{m_1}
\times\bepsilon_{m_1^\prime}^*)
\cdot(\bepsilon_{m_2}\times\bepsilon_{m_2^\prime}^*)\nonumber\\
&\approx&-0.4958{N_c{g^2}\omega_0\over16\pi}\delta_\alpha^{\alpha^\prime}
\delta_{\beta^\prime}^\beta(\bepsilon_{m_1}
\times\bepsilon_{m_1^\prime}^*)
\cdot(\bepsilon_{m_2}\times\bepsilon_{m_2^\prime}^*).
\end{eqnarray}
For comparison, we put in the numbers for the quartic contribution:
\begin{eqnarray}
\bra{m_1^\prime, m_2^\prime}\Delta_{quartic}\ket{m_1,m_2}
&\approx&0.1129{N_cg^2\omega_0\over16\pi}\nonumber\\
&&[(\bepsilon_{m_1}\times\bepsilon_{m_1^\prime}^*)
\cdot(\bepsilon_{m_2}\times\bepsilon_{m_2^\prime}^*)
+(\bepsilon_{m_1}\times\bepsilon_{m_2})
\cdot(\bepsilon_{m_1^\prime}^*\times\bepsilon_{m_2^\prime}^*)]
\end{eqnarray}
Combining the quartic and exchange contributions gives the
level pattern (0.3829, $-$0.1571, $-$.7658) in units of
$N_cg^2\omega_0/16\pi$, restoring an anti-ferromagnetic
coupling.

It remains to evaluate the annihilation graph. In this case the
currents carry the time dependence $e^{\pm2i\omega_0 t}$, so
the propagator is no longer the static Green function. For this
graph the currents turn out to be
\begin{equation}
{\cal J}^1_k=-{9K^2\over4\pi\omega_0^2}
{j_1(\omega_0r)^2\over2\omega_0 r^2}[{\bf r}\times(\bepsilon_{m_1}
\times\bepsilon_{m_2})]_k\qquad{\cal J}^2_k=e^{2i\omega_0 t}
{9K^2\over4\pi\omega_0^2}
{j_1(\omega_0r)^2\over2\omega_0 r^2}[{\bf r}\times(\bepsilon_{m_1^\prime}^*
\times\bepsilon_{m_2^\prime}^*)]_k,
\end{equation}
and 
$${\cal A}_k({\bf x})\equiv
\int d{\bf y}\int dt e^{2i\omega_0 t}D_{kl}({\bf x},t;{\bf y},0)
{\cal J}^1_k({\bf y})
$$
satisfies the inhomogeneous Helmholtz equation 
\begin{equation}
(-\nabla^2-\omega^2){\cal A}_l={\cal J}^1_l=-{9K^2\over4\pi\omega_0^2}
{j_1(\omega_0r)^2\over2\omega_0 r^2}[{\bf r}\times(\bepsilon_{m_1}
\times\bepsilon_{m_1^\prime}^*)]_l,
\end{equation}
with $\omega=2\omega_0$, but it is clearer to write the solution for
general $\omega$.
Again with the {\it ansatz} ${\cal A}_l=A(r)[{\bf r}\times(\bepsilon_{m_1}
\times\bepsilon_{m_2})]_l$, this becomes
\begin{eqnarray}
A^{\prime\prime}+{4\over r}A^\prime+\omega^2 A={9K^2\over4\pi\omega_0^2}
{j_1(\omega_0r)^2\over2\omega_0 r^2},
\end{eqnarray}
with the boundary condition remaining $RA^\prime(R)+2A(R)=0$. Putting these
into the formula for the contribution to the energy shift gives
\begin{eqnarray}
\bra{f}\Delta E\ket{i}_{annih}
= {N_c{g^2}}\delta_\alpha^{\alpha^\prime}
\delta_{\beta^\prime}^\beta {3K^2\over\omega_0^3}\int_0^R r^2dr
A(r)j_1(\omega_0r)^2(\bepsilon_{m_1}\times\bepsilon_{m_2})
\cdot(\bepsilon_{m_1^\prime}^*\times\bepsilon_{m_2^\prime}^*).
\end{eqnarray}
We see immediately that this graph only contributes to the shift
of the triplet spin 1 state. This is, of course, not surprising because it
describes a direct mixing between two gluon and one gluon states
and the one gluon state is pure spin 1.

The solution for $A$ with the correct boundary conditions is easily
found to be, for general $\omega$,
\begin{eqnarray}
A(r)={9K^2\over4\pi\omega_0^2r}{\omega\over2\omega_0}
\int_0^R r^\prime dr^\prime
j_1(\omega r_<)\left[y_1(\omega r_>)-
{\omega Ry_0(\omega R)-y_1(\omega R)\over \omega Rj_0(\omega R)
-j_1(\omega R)}j_1(\omega r_>)\right]
j_1(\omega_0 r^\prime)^2,
\end{eqnarray}
where we use the standard notation $r_< (r_>)$ for the lesser(greater) 
of the pair $r, r^\prime$. We want $\omega=2\omega_0$, but 
with $\omega$ general we 
should recover the previous case for $\omega\to 0$, 
a useful check on numerical work. Stripping off the color factors,
the annihilation graph contribution to the level shift becomes
\begin{eqnarray}
&&\bra{f}\Delta_{annih}\ket{i}
= {N_cg^2\omega_0\over16\pi}{27\eta\over\chi(z_0)^2}
\int_0^{z_0} zdzj_1(z)^2\left[y_1(\eta z)-
{\eta z_0y_0(\eta z_0)-y_1(\eta z_0)\over \eta z_0j_0(\eta z_0)
-j_1(\eta z_0)}j_1(\eta z)\right]\nonumber\\
&&\qquad\qquad\qquad\int_0^z z^\prime dz^\prime j_1(\eta z^\prime)
j_1(z^\prime)^2(\bepsilon_{m_1}\times\bepsilon_{m_2})
\cdot(\bepsilon_{m_1^\prime}^*\times\bepsilon_{m_2^\prime}^*).
\end{eqnarray}
where we have defined $\eta\equiv\omega/\omega_0\to2$ for this
graph. For this value of $\eta$, the numerical evaluation of
the above expression gives
\begin{eqnarray}
&&\bra{f}\Delta_{annih}\ket{i}
\approx 0.1430{N_cg^2\omega_0\over16\pi}
(\bepsilon_{m_1}\times\bepsilon_{m_2})
\cdot(\bepsilon_{m_1^\prime}^*\times\bepsilon_{m_2^\prime}^*).
\end{eqnarray}
Finally putting all contributions together we arrive at the total
level shift matrix
\begin{eqnarray}
&&\bra{m_1^\prime, m_2^\prime}\Delta_{total}\ket{m_1,m_2}
\approx\nonumber\\
&&\qquad{N_cg^2\omega_0\over16\pi}[-0.3829(\bepsilon_{m_1}
\times\bepsilon_{m_1^\prime}^*)
\cdot(\bepsilon_{m_2}\times\bepsilon_{m_2^\prime}^*)
+0.2559(\bepsilon_{m_1}\times\bepsilon_{m_2})
\cdot(\bepsilon_{m_1^\prime}^*\times\bepsilon_{m_2^\prime}^*)],
\end{eqnarray}
which translates to the level pattern $(0.3829,0.1289,-0.7658)$
in units of $N_cg^2\omega_0/16\pi$ for spins $2, 1, 0$ respectively.

The upshot of our calculations is that the perturbative interactions
between nearest neighbors on the gluonic chain described by
't Hooft's large $N_c$ limit have the correct sign and spin
dependence to provide a satisfactory pattern of single glueball
states. But of course perturbation theory is only valid for
weak coupling (at short distances), and a near neighbor bond of massless
gluons requires this coupling to be at least of order 1. 
As the coupling increases at larger distances, perturbation
theory will break down, but it is perhaps not too much to hope
that the qualitative pattern of interactions is not drastically
changed. The anti-ferromagnetic character of the spin-spin
interactions is very reassuring. Given that the bonds
do form, even a relatively weak residual anti-ferromagnetic
interaction should be enough to guarantee that the 
lowest excitations of a long chain have low spin. 

It is noteworthy that this pattern would {\it not} hold if the
only interaction were due to the $A^4$ term: gluon exchange
is essential. Thus, for example, the fishnet diagrams,
based on a quartic coupling and
analyzed in the large coupling light-front studies of \cite{browergt},
{\it cannot} contain the essential physics of large $N_c$
QCD. Those diagrams were singled out by setting up a cutoff 
model in which the strong coupling limit ``froze'' the
$P^+$ distribution to be uniform amongst all the gluons in
the chain. In fact, this frozen $P^+$ distribution led to a string
picture of the gluonic chain with the degrees of freedom of a critical
string, whereas QCD must be subcritical (4$<$26!).
To bring into a strong coupling expansion the exchange effects necessary
for anti-ferromagnetic interactions, a more sophisticated
cutoff model which allows fluctuations in the $P^+$
distribution at large coupling must be devised. It is from these fluctuations
that we can expect the Liouville world sheet field, necessary for
subcritical string theory, to emerge. As described in \cite{gubserkp}
the Liouville field should also be the ``5th'' dimension in any
alleged QCD/5dString connection. 

\medskip
\noindent \underline{Acknowledgements}. I would like to thank Joel
Rozowsky and Klaus Bering for helpful comments on the manuscript.

\bibliography{../larefs,../textbooks}
\bibliographystyle{unsrt}

%
\end{document}